# Neutral polyphosphocholine-modified liposomes as boundary superlubricants

Weifeng Lin, Nir Kampf, and Jacob Klein*

Department of Molecular Chemistry and Materials Science, Weizmann Institute of Science, Rehovot 76100, Israel

* jacob.klein@weizmann.ac.il

## Abstract

Boundary lubrication is associated with two sliding molecularly thin lubricated film-coated surfaces, where the energy dissipation occurs at the slip-plane between lubricated films. The hydration lubrication paradigm, which accounts for ultralow friction in aqueous media, has been extended to various systems, with phosphatidylcholine (PC) lipids recognized as extremely efficient lubrication elements due to their high hydration level. In this work, we extend a previous study (Lin et al., Langmuir 35 (2019) 6048-6054), where a charged lipid- poly(2-methacryloyloxyethyl phosphorylcholine) (PMPC) conjugate was prepared, to the very different case of  a neutral lipid-PMPC) conjugate. This neutral molecule stabilizes the liposomes by attaching highly water-soluble PMPC to the surface of liposomes with its lipid moieties incorporated in the lipid bilayers. Such neutral polyphosphocholinated liposomes provide a surface lubricity which is well within the superlubrication regime ($\mu \approx 10^{-3}$ or even lower). In contrast, negatively charged lipid/polyphosphocholine conjugates modified liposomes were unable to adsorb on negatively-charged (mica) surfaces. Our method provides stable liposomes that can adsorb on negatively charged surfaces and provide superlubricity.

# 1. Introduction

Phosphatidylcholine (PC) lipids, consisting of one hydrophilic group and two hydrophobic tails, present as a major component of biological membranes.[1] As surface-active compounds with highly-hydrated headgroups, PC lipids play essential roles in various interfacial biological activities.[2-4] They are ubiquitous in the synovial fluid and on the most superficial layer of articular cartilage,[5, 6] contributing to the ultralow friction (friction coefficient of $\mu$ = friction force/normal load ≈ $10^{-3}$, up to physiologically high pressures[7]).[8] The low friction provided by the PC lipids has been extensively examined using nanotribology measurements (such as surface force balance, SFB, and friction force microscopy),[9-12] as well as macroscopic tribometry measurements.[13-18] The origin of such low friction is due to the hydration lubrication mechanism.[19, 20] The highly-hydrated phosphocholine headgroups can sustain large pressures without the hydration water being squeezed out; at the same time, the hydration shell can exchange rapidly with free water molecules to ensure a fluid response during sliding.[21] This combination leads to extreme reduction of frictional energy dissipation (in other words, ultralow shear stresses) while surfaces slide past each other under high normal stresses.

Phosphocholine-exposing vesicles (liposomes) provide an effective method of providing superior lubricity to interfaces, and have been proposed to relieve pathologies (such as osteoarthritis or dry eye syndrome) associated with lubricating failure.[8] As potential drug carriers, liposomes are able to load simultaneously hydrophilic drugs into the aqueous lumen and hydrophobic drugs into the lipid bilayer membranes.[22] However, such liposomes without stabilizer have the tendency to aggregate (or fuse) over time. Therefore, these liposomes must be stabilized for keeping their integrity (for example during storage or delivery) and preventing such aggregation.[23, 24] To overcome the above deficiency, sterically stable liposomes were made, containing natural or synthetic polymer-modified lipid derivatives on the liposomes' surface.[25, 26] The types of polymers include: polyethylene glycol (PEG),[27] polyzwitterions,[28, 29] poly[N-(2-hydroxypropyl) methacrylamide] (PHPMA),[30] poly(2-oxazoline),[31] poly(N-vinyl-2-pyrrolidone) (PVP),[32] hyaluronic acid (HA),[33] among which PEGylation is the most widely used.[34, 35] However, water molecules form relatively weak hydrogen bonding with PEG moieties,[36] which may impair the efficient lubricating properties provided by PC vesicles. Zwitterionic polymers are a class of polymers having equal cationic and anionic groups along their polymer chains, with much stronger hydration due to their ionic solvation nature compared with PEG.[37, 38] Zwitterionic poly[2-(methacryloyloxy)ethyl phosphorylcholine] (PMPC) is distinguished among the zwitterionic materials,[39-42] being a methacrylate with a highly-hydrated phospholipid polar group in the side chain, essentially a phosphocholine-like monomer structure resembling the headgroup structure of naturally-occurring PC lipids. In our previous study,[43] PMPC-functionalized (PMPCylated) liposomes were prepared, which were stable against aggregation in aqueous media. Furthermore, the PMPCylated liposomes adsorbed onto mica surfaces (at physiological salt concentrations or low pH, which screens or neutralizes their negative charge), forming boundary layers providing excellent lubricity under physiological ionic conditions. The friction coefficient of such a system is equivalent to the non-functionalized PC liposomes with an order of magnitude lower friction coefficient than for PEGylated liposomes. It was also found that such PMPCylated liposomes intra-articularly (IA) injected into mice joints showed a significant increase in retention half-life compared with PEGylated liposomes.[44]

In order to reduce friction between surfaces, the liposomes must adsorb on the substrate to form boundary layers. Non-functionalized PC liposomes adsorb on negatively charged surfaces via dipole-charge interactions between the phosphocholine zwitterions and the negative surface charges. Liposomes stabilized by the lipid-PMPC conjugate (distearoylphosphatidylethanolamine-PMPC, DSPE-PMPC) even at a rare concentrations as low as 1.5%, have a net negative charge and can adsorb on negatively charged surfaces (such as a mica surface), only by charge screening under 0.15 M salt solution.[43] Molecularly-smooth mica surfaces are chosen in this study, which are clearly different to articular cartilage (although both are negatively charged), but they allow good control over surface roughness in order to provide molecular insight into the mechanisms of boundary friction which occurs at the slip plane and is largely independent of the substrate. On the other hand, lots of studies related to boundary lubricants don't pay attention to the stability of the lubricants.[11-13] The current study aims to overcome this limitation and make up the overlook, stabilized liposomes with zero-charge was prepared by incorporating a newly synthesized distearoylglycerol-PMPC (DSG-PMPC) neutral lipid-PMPC conjugate into the PC-liposomes. We investigate the stability of DSG-PMPC-modified liposomes by dynamic light scattering (DLS), and the adsorption of these liposomes in pure water on mica by atomic force microscopy (AFM). After ensuring the adsorption of the neutral stabilized liposomes onto the mica, we examine their lubrication ability by measuring the normal and shear forces using a surface force balance (SFB) technique.

## 2. Experimental

### 2.1 Materials

1,2-distearoyl-sn-glycerol (DSG, ≥98%) was purchased from Bachem (Bubendorf, Switzerland). Hydrogenated soy phosphatidylcholine (HSPC, >99%) was purchased from Lipoid (Ludwigshafen, Germany). α-bromoisobutyryl bromide (BIBB, 98%), triethylamine (≥99.5%), copper(I) bromide (CuBr, 99.999% trace metals basis), *N,N′,N′,N″,N″*-pentamethyldiethylenetriamine (PMDETA, 99%), dichloromethane (anhydrous, ≥99.8%), chloroform (anhydrous, ≥99%), ethyl ether (anhydrous), and silver beads (99.9999%) were purchased from Sigma-Aldrich (Israel). Hydrochloric acid (37%), hydrogen peroxide ($H_2O_2$, 30%), and sulfuric acid ($H_2SO_4$, 95-98%) were purchased from Bio-Lab (Jerusalem, Israel). Nitric acid ($HNO_3$, 65%) was purchased from Merck KGaA (Darmstadt, Germany). Ethanol (99.5%, HPLC grade) was purchased from J.T Baker (Poland, USA). Ruby muscovite mica (high grade) was purchased from S&J Trading (New York, USA). 2-(methacryloyloxy)ethyl phosphorylcholine (MPC) was obtained from Vertellus Biomaterials (Basingstoke, U.K.). Water was highly purified by a Barnstead Nanopure Diamond water system and had a resistivity of 18.2 MΩ cm at 25 °C with a total organic content less than 1 ppb. Ethanol was filtered through a polyethersulfone membrane (0.45 μm) from a pressure rinser (Pall Corporation, USA). Prior to any experiment, the glassware was cleaned in piranha solution (2:1 mixture of $H_2SO_4$ and 30% $H_2O_2$), washed with excess of purified water, and sonicated in purified water and ethanol for 10 min, respectively. Stainless steel tools were passivated in 30% aqueous $HNO_3$ (60 °C) for 0.5 hour, followed by sonication in purified water and ethanol for 10 min, respectively.

## 2.2 Synthesis of DSG-PMPC

Neutral lipid-PMPC conjugate (DSG-PMPC) was synthesized by atom transfer radical polymerization (ATRP)[45, 46] using DSG-Br as the initiator (as shown in Scheme 1).

Synthesis of DSG-Br initiator: triethylamine (0.5 mL, 3.92 mmol) and DSG (1.53 g, 1.95 mmol) were added to 100 mL of anhydrous chloroform, and the mixture was stirred at 40 °C for 0.5 hour. When BIBB (0.192 mL, 1.95 mmol) was added dropwise to a mixture of these two compounds, the mixture became clear and was stirred overnight. The solution was then washed with purified water and 1 M hydrochloric acid solution three times, respectively. A white powder was obtained after solvent removal by rotary evaporation (yield: 95%). $^1$H NMR (300 MHz, CDCl$_3$): $\delta$ = 5.34 (–OCH(CH$_2$O)$_2$–); $\delta$ = 4.32 (–OCH(CH$_2$O)$_2$–); $\delta$ = 2.34 (–COCH$_2$CH$_2$–); $\delta$ = 1.86 (–COC(Br)(CH$_3$)$_2$); $\delta$ = 1.63 (–COCH$_2$CH$_2$–); $\delta$ = 1.27 (–(CH$_2$)$_{14}$–); $\delta$ = 0.89 (–CH$_2$CH$_3$).

Synthesis of DSG-PMPC by atom-transfer radical polymerization (ATRP): DSG-Br (141 mg, 0.15 mmol), MPC (800 mg, 2.7 mmol), and CuBr (21 mg, 0.15 mmol) were dissolved in solvent mixtures (5 mL dichloromethane and 5 mL ethanol). The mixture was kept under N$_2$ flow for 30 minutes to eliminate the presence of oxygen. PMDETA (60 µl, 0.3 mmol) was then quickly injected. The mixture was stirred for 12 hours at room temperature. Precipitate was collected by adding the mixture to excess ethyl ether. The precipitate was dissolved in ethanol and dialyzed against purified water for 2 days (CelluSep membrane, MWCO 1000). The lipid-polymer conjugate DSG-PMPC powder was obtained after lyophilization (yield: 67%). $^1$H NMR (300 MHz, CDCl$_3$/CD$_3$OD 1:1): $\delta$ = 0.32−0.77 (CH$_3$−CR−CH$_2$−), $\delta$ = 0.77−1.05 (−(CH$_2$)$_{14}$−CH$_3$), $\delta$ =

1.15−1.85 (−CH$_2$−CR−CH$_3$), δ = 2.972 (−N$^+$(CH$_3$)$_3$), δ = 3.37 (−CH$_2$−N$^+$(CH$_3$)$_3$), δ = 3.69 (−CH$_2$−PO$_4$$^-$−), δ = 3.84 (−CO−CH$_2$−), δ = 3.94 (−PO$_4$$^-$−CH$_2$−).

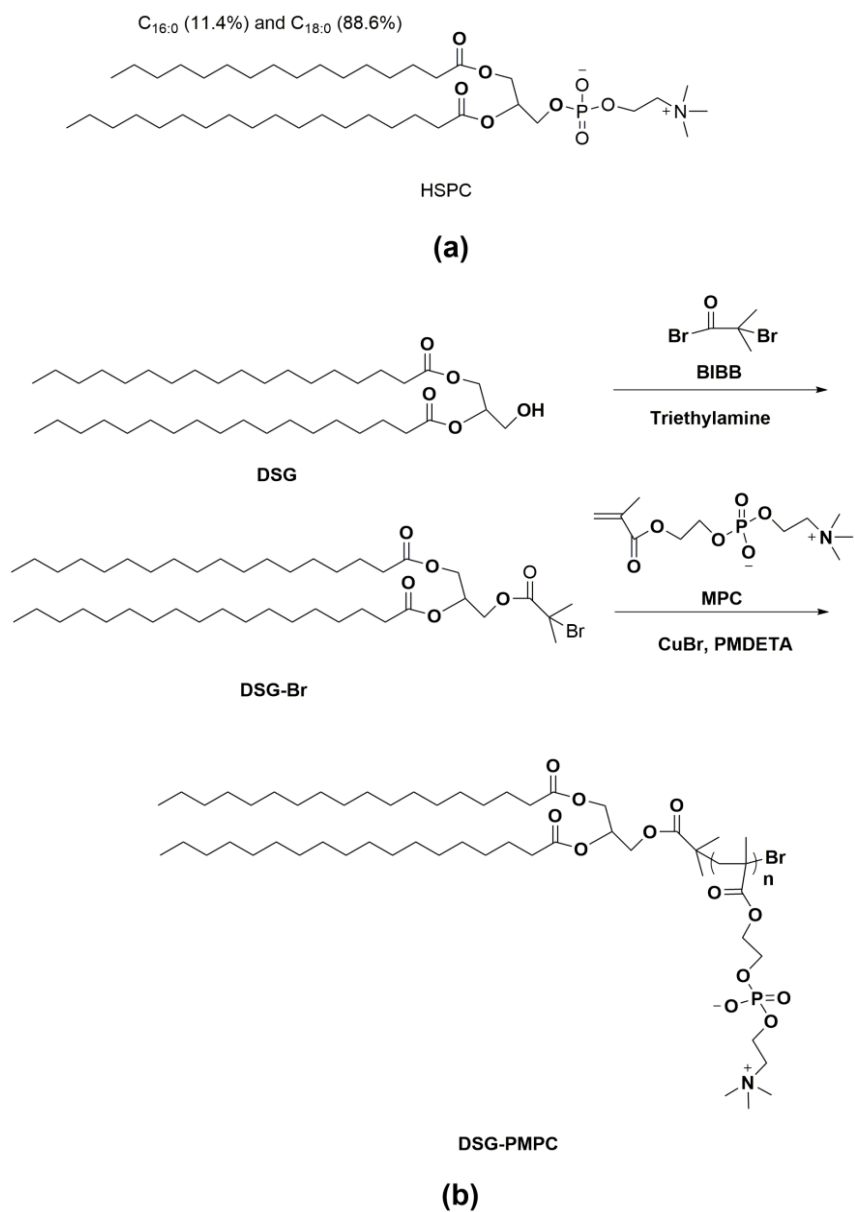

**Scheme 1.** (a) Chemical structure of HSPC lipid; (b) Synthesis procedure of the ATRP initiator DSG-Br and neutral lipid-polymer conjugate DSG-PMPC.

## 2.3 Liposome Preparation

Small unilamellar vesicles (SUVs) with mixed lipid composition were made by a thin-Film hydration followed by extrusion method.[43, 47] Pure HSPC (118 mg) powder or HSPC (115 mg)/DSG-PMPC (11 mg) mixture was dissolved in methanol (1 mL) and chloroform (1 mL). The organic solvent was evaporated by blowing a stream of dry nitrogen gas followed by vacuum desiccation overnight. The lipid mixture was suspended in pure water and sonicated for 15 mins at 70 °C (15 °C above the transition temperature) followed by 5 mins vortex to yield multilamellar vesicles (MLVs). SUVs were obtained (∼70 nm in diameter, 15 mM) by downsizing the MLVs with stepwise extrusion (Northern Lipids, Burnaby, Canada) through polycarbonate membranes having pore sizes with 400 nm (5 cycles), 100 nm (8 cycles), and 50 nm (12 cycles).

## 2.4 Dynamic Light Scattering (DLS)

Size distribution and zeta potential of the liposomes (1 mM in pure water) were measured on a Malvern Zetasizer Nano ZSP instrument with a red laser (633 nm) and a scattering angle of 173° (performed at 25.0 °C). The hydrodynamic diameter, $D_h$, of spherical liposomes was calculated from the Stokes−Einstein equation $D_h = k_B T/3\pi\eta D_0$, where $D_0$ is the measured diffusion coefficients, $k_B$ is the Boltzmann constant ($1.38064852 \times 10^{-23}$ J/K), $T$ is the temperature, and $\eta$ is the viscosity. Zeta potential ($\zeta$-potential) was calculated automatically from the electrophoretic mobility ($v_E$) of liposomes based on the Hückel equation $v_E = 2\pi\varepsilon_0\varepsilon_r\zeta/3\eta$, where $\varepsilon_r$ is the dielectric constant, $\varepsilon_0$ is the electrical permittivity of a vacuum, and $\eta$ is the viscosity.

## 2.5 Atomic Force Microscopy (AFM)

Samples were prepared by adding a 0.3 mM liposome dispersion to a Petri dish (60 mm × 15 mm) with a freshly cleaved piece of mica (ca. 4 cm$^2$) glued to the inner bottom surface. After overnight incubation, the excess liposomes were infinitely diluted to avoid passing the air-water interface. An Asylum MFP-3D$^{TM}$ stand-alone AFM (Oxford Instruments, Santa Barbara, USA) was used for surface-imaging of the liposomes on the mica substrates. The surfaces were scanned in non-contact mode under pure water using a silicon probe (SNL-10, Bruker) with V-shaped cantilever having a nominal spring constant of 0.35 N/m. The AFM tip holder was irradiated in a ProCleaner$^{TM}$ Plus UV-ozone cleaning (BioForce Nanosciences, Utah, USA) for 20 min prior to use in order to remove molecular levels of contamination.

## 2.6 Surface Force Balance (SFB)

An SFB was used to directly measure the normal ($F_n$) and shear ($F_s$) forces as a function of the absolute surface separation $D$, which is in contrast with scanning probe methods (including colloidal probe AFM) where only relative motions of the surfaces can be measured. The working principles and detailed experimental procedures of SFB technique have been described previously.[48] Briefly, two back-silvered atomically-smooth mica sheets (area of ∼1 × 1 cm$^2$ and thickness of ∼2.5-4.0 μm) were glued onto plano-cylindrical glass lenses (radius of 1 cm) and mounted in a cross-cylinder configuration. The absolute surface separation ($D$, ±0.3 nm) was measured from the wavelength of optical interference fringes of equal chromatic order (FECO). The lower surface is mounted onto a horizontal stainless steel leaf spring (spring constant $K_n$ = 127 N/m) and its bending, Δ$D$, is measured via the interferometry technique to determine the normal force $F_n(D) = K_n\Delta D$. In the Derjaguin approximation,[49] $F_n(D)/R$ is plotted to relate the normal force $F_n$ with the interaction energy per unit area between flat surfaces. The upper surface mount is coupled to a four-sectored piezoelectric tube (PZT), supplied by Ferroperm (Pz29). By applying suitable potentials to opposite sectors, lateral motion $\Delta x_0(t)$ is obtained with respect to the lower surface. Shear forces between the surfaces $F_s(D)$ are transmitted to vertical springs ($K_s$ = 300 N/m), whose bending as a function of time $\Delta x(t)$ is measured, via an air-gap capacitor, and yields the shear forces $F_s(D, t) = K_s\Delta x(t)$. The mean pressure ($P$) between two surfaces can be calculated from the contact area ($A$) as $P = F_n/A = F_n/\pi a^2$, where $a$ is the radius of contact area. At high loads, $a$ can be directly measured from the flattened fringe image. At low loads, where unclear fringe flattening is seen, $a$ can be evaluated from Hertzian contact mechanics,[50] $a = (F_n R/K)^{1/3}$ where K is the known effective elastic modulus. Data shown are based on three independent experiments, with a number of contact points in each experiment.

## 3. Results and discussion

The ATRP initiator (DSG-Br) was obtained through the esterification reaction of the terminal hydroxyl group of the DSG with 2-bromoisobutyryl bromide. The $^1$H NMR spectrum of DSG-Br (Figure 1a) shows a characteristic signal at $\delta$ = 1.86 (–COC(Br)(CH$_3$)$_2$). DSG-PMPC synthesis was carried out via ATRP using DSG-Br as the initiator (Scheme 1), and the characteristic peaks of the DSG-Br initiator ($\delta$ = 0.90, –(CH$_2$)$_{14}$–) and MPC monomer ($\delta$ = 3.37, −CH$_2$−N$^+$(CH$_3$)$_3$) determined that the degree of polymerization of DSG-PMPC is ca. 14 (molecular weight of PMPC is ca. 4.2 kDa), as shown in Figure 1b. The DSG-PMPC neutral lipid-polymer conjugate, was incorporated into the HSPC-SUVs by its hydrophobic distearoyl part, forming a PMPC-stabilized liposomes (Figure 2a). HSPC is a mixture of dipalmitoylphosphatidylcholine (DPPC) and distearoylphosphatidylcholine (DSPC) lipids (average fatty acid distribution of $C_{16:0}$ (11.4%) and $C_{18:0}$ (88.6%)), and was chosen due to the following reasons). 1) As our study is aimed to be involved in OA treatments, these two lipids are present in the synovial fluid and on cartilage surfaces.[5, 6] 2) Compared with DSPC and DPPC, HSPC is practically preferred because hydrogenating extracted soybean PC (natural lipids) is easier than synthesizing the whole lipids.[51] 3) HSPC has been demonstrated as an excellent boundary lubricant [10] and is an important lipid-based carrier for drug loading and sustained release.[52, 53] The neutral polyphosphocholine-modified SUVs were made by mixing the DSG-PMPC molecules (2%) with the HSPC lipid in pure water followed by film hydration−extrusion to form uniform ca. 70 nm vesicles (see Methods).

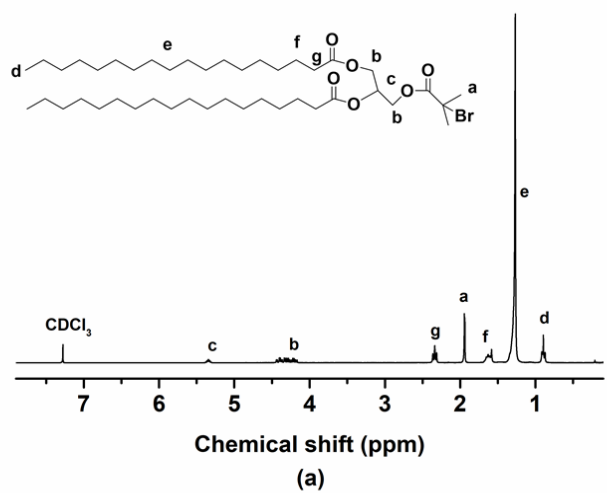

(a)

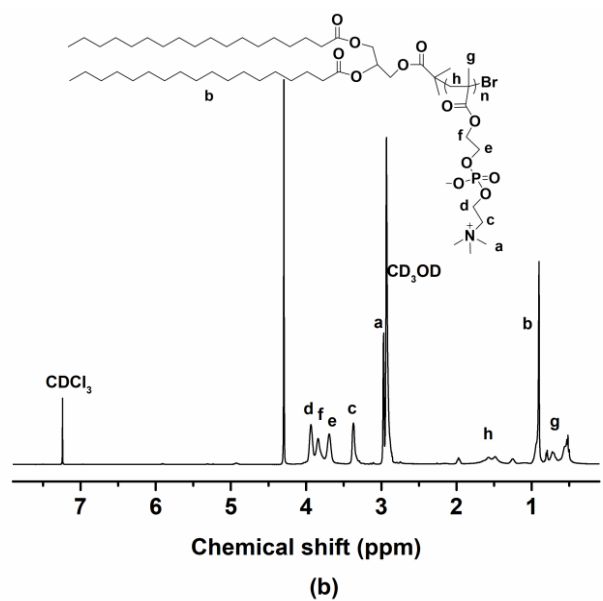

**Figure 1**. $^1$H NMR characterization of (a) ATRP initiator DSG-Br (in CDCl$_3$) and (b) neutral lipid-PMPC conjugate DSG-PMPC (CD$_3$OD/CDCl$_3$ = 1/1, v/v).

The stability of the two types of liposomes (HSPC-SUV and HSPC/DSG-PMPC-SUV) was determined by dynamic light scattering (DLS), as shown in Figure 2b. The results clearly show the aggregation process within 7 days of the HSPC-SUVs with an increasing average size from ca. $D_h$ = 70 nm (monodispersed with PDI < 0.1) to two peaks with a second peak at $D_h$ = 500-600 nm (PDI > 0.3). In contrast, HSPC/DSG-PMPC-SUVs remain stable (unchanged in size or PDI) for at least 1 month. It is worth to mention that the ester linkage on the DSG-PMPC conjugate could be less hydrolytically stable than the amide linkage on the DSPE-PMPC conjugate. However, in practice such stability was maintained at least for one month (measured by DLS). In our further work, the effect of linkage (ester or amide) on the long-term stability of the liposomes would be discussed. Zeta ($\zeta$) potential measurements on stable liposomes showed a neutral potential (-0.1 ± 0.4 mV) in pure water, while $\zeta$ (HSPC-SUV) = 1.2 ± 0.9 mV and $\zeta$ (HSPC/DSPE-PMPC-SUV = -25.2 ± 3.2 mV). These results support the fact that the stability of HSPC/DSG-PMPC-SUV is due to the steric interactions (including steric hydration forces) of the polymer chains rather than electrostatic repulsive effects.

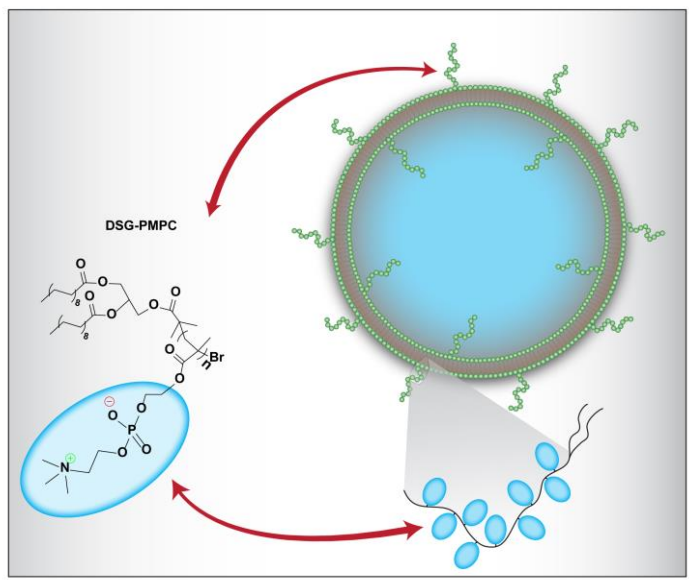

(a)

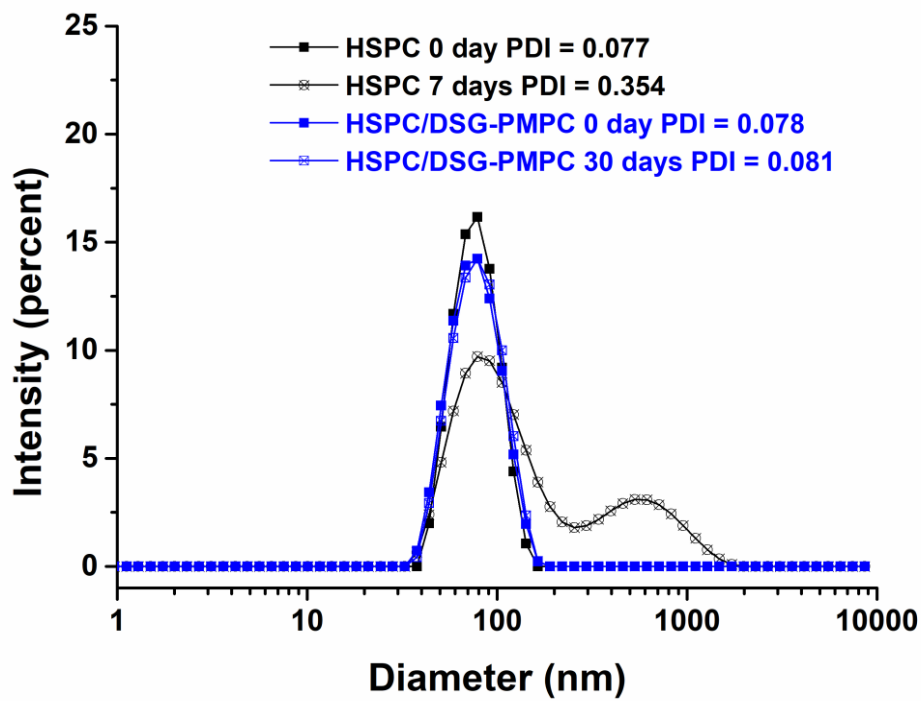

(b)

**Figure 2**. (a) Schematic of the neutral lipid-PMPC conjugate and its incorporation in a PC liposome. (b) Size distribution of HSPC-SUVs (polydispersity, PDI), either unfunctionalized or stabilized by DSG-PMPC, in pure water at different times following preparation. Measurements were performed at 1 mM lipid concentration.

The mica substrates were incubated overnight in the respective liposomal dispersions, which then were infinitely diluted with purified water to remove excess liposomes. Tapping-mode AFM was used to characterize lipid surface morphology after such incubations and infinite dilutions as shown in Figure 3. AFM micrographs demonstrate an adsorbed, closely packed, DSG-PMPC-modified vesicles on the mica surface with some extra vesicles in an overlayer (Figure 3b). In contrast, no DSPE-PMPC-modified vesicles were observed on the mica surface after incubation and dilution to pure water (Figure 3a; similar micrographs indicating no liposome adsorption were obtained when the AFM imaging was carried out following incubation under dispersion with no further dilution).

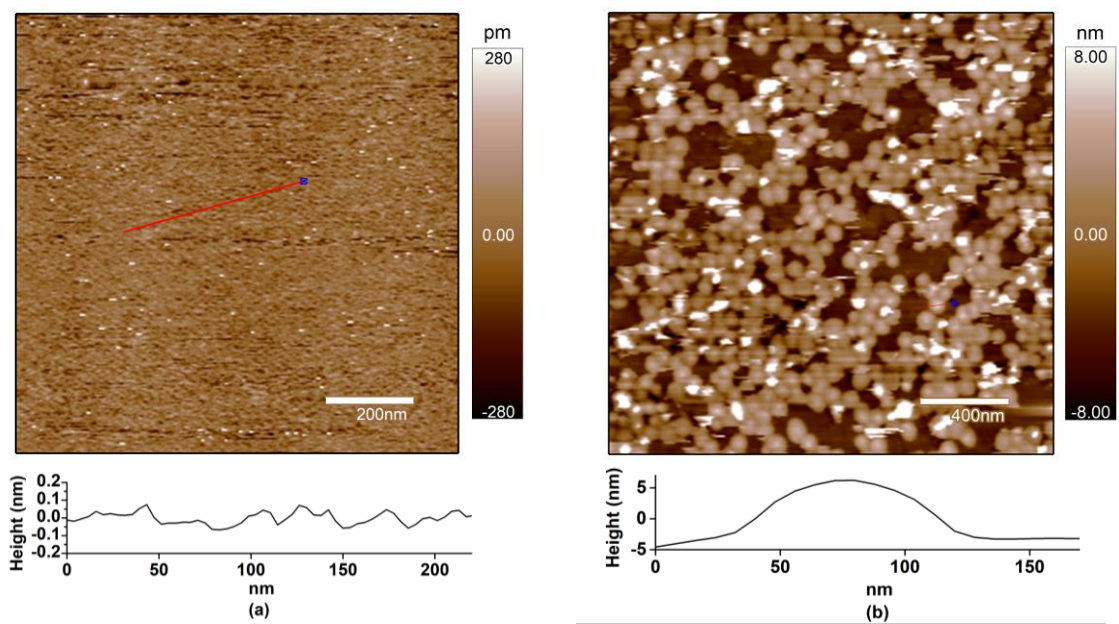

**Figure 3.** Non-contact mode AFM images and cross section of mica after overnight incubation in (a) HSPC/DSPE-PMPC-SUV or (b) HSPC/DSG-PMPC-SUV, followed by infinite dilution with pure water. Similar images were obtained at least in three different points on the substrates.

Normal force $F_n(D)$ versus surface-separation $D$ profiles and friction force $F_s$ versus $F_n$ profiles between mica surfaces coated with DSG-PMPC-modified liposomes are shown in Figure 4. As the two liposome-coated surfaces came closer from ca. 500 nm, interaction was undetected above the scatter at surface separations $D$ around ca. 200 nm (Figure 4a). The range, exponential increase of the forces with decreasing $D$, and decay length of the interaction suggests that initially (50 – 100 nm < $D$ < ca. 200 nm) it is likely to be a double-layer electrostatic (DLVO) repulsion, since the underlying mica substrates remain charged even after the adsorption of the neutral DSG-PMPC liposomes. At the same time, there may also be a steric component at longer range due to loosely-attached liposomes (overlaying the surface-adsorbed vesicles, as suggested by Figure 3b, and seen also in earlier studies,[10] though these are likely removed on closer approach. As $D$ decreases, the steric repulsion due to the adsorbed liposomes begins to dominate (compare DLVO predicted orange curve with the force data), until for $D$ < ca. 50 nm the interactions, both normal and eventually frictional forces, are completely dominated by the steric forces at high compressions. At higher compressions, at say $D$ < ca. 70 nm, where normal forces increase more rapidly than exponentially with decreasing $D$, the steric repulsion due to the distortion of one layer of absorbed liposome becomes dominant. Dimensions of the PMPC coils could be estimated from the study by Kobayashi et al.[54], where the measured hydrodynamic radius $R_H$ for a free PMPC chain of $M_n$ = 234 K was 11.5 nm. Taking $R_H$ as close to the coil size $R$ in good solvent and $R \approx M^{3/5}$ yields the estimate $R_{PMPC} \approx 1.0$ nm ($M \approx 4200$ in this case, while the value of $R$ might well be underestimated due to the shortness of the PMPC moieties). The interanchor spacing of PMPC chains $s$ is given by $s \approx (A_{HSPC}/X_{DSG\text{-}PMPC})^{1/2} = 6.5$ nm, where $A_{HSPC}$ is the area per HSPC molecule (0.64 nm$^2$) and $X_{DSG-PMPC}$ is mole fraction of DSG-PMPC (1.5 mol%), which is considerably larger than the coil size $2R$ (non-interacting polymer coils) and is considered as in the mushroom region. Upon strong compression, 'hard wall' separations of 21 ± 2 nm were obtained, corresponding respectively to 2 compressed liposomes layers plus the thickness of PMPC moieties (the thickness of one bilayer and one flattened SUV layer is ca. 5 nm and ca. 10 nm). The hard wall repulsion at $D \approx 20$ nm is characteristic of two strongly-compressed liposomes, indicating that each liposome layer is attached directly to the mica surface with no underlying bilayer.

The frictional forces $F_s$ transmitted between the interacting surfaces are measured from the corresponding shear traces. The associated friction coefficients are in line with values that have been previously observed between mica sheets bearing case of HSPC-SUVs.[10] $\mu = 0.0008\text{-}0.002$ (Figure 4b) was found for the two mica surfaces aliasing past each other across the DSG-PMPC-modified liposomes layers (up to mean contact pressures ca. 92 atm). In both cases, these low friction coefficients are within the superlubrication regime and are attributed to the hydration lubrication mechanism of the exposed, highly hydrated phosphocholine headgroups of the PMPC moieties and PC lipids.

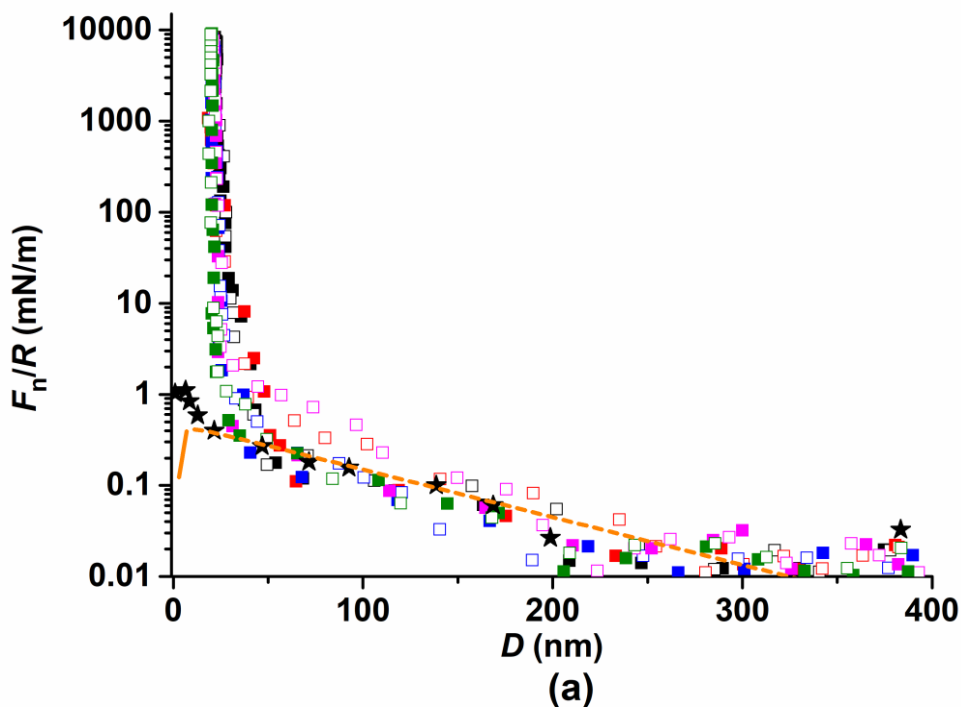

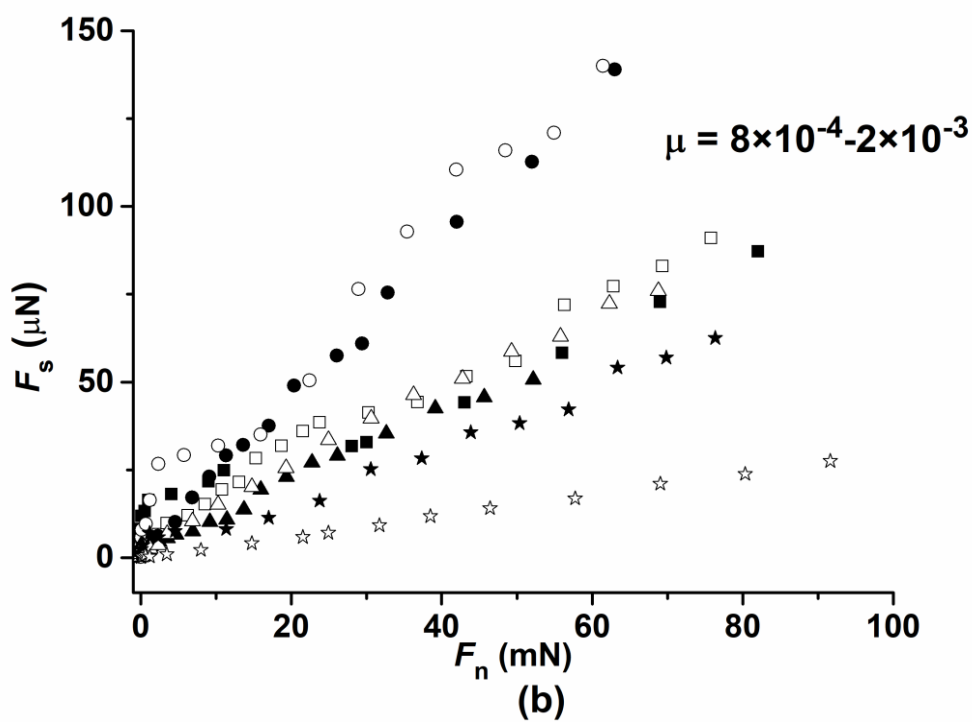

**Figure 4.** (a) Normalized force profiles $F_n(D)/R$ between DSG-PMPC-modified liposomes layers in pure water. $R$ is the mean radius of curvature of the mica sheets. Different shaped symbols (except for the star symbols corresponding to the data of two bare mica surfaces across pure water) correspond to different contact positions with first approaches (filled symbols) and second

approaches (empty symbols) with DSG-PMPC-modified liposomes layers. The orange dashed line is the theoretical Derjaguin–Landau–Verwey–Overbeek DLVO fitting of the $F_n(D)/R$ vs $D$ results, corresponding to Debye lengths ($\kappa^{-1}$) of 83 nm. (b) Frictional forces $F_s$ versus $F_n$ between DSG-PMPC-modified liposomes layers, where the $\mu$ range given for each system is based on the extremal data points.

## 4. Conclusions

PMPCylated PC liposomes represent a class of liposomes that form a boundary layer that lubricates much more efficiently than PEGylated liposomes (where PEGylation is most commonly used for liposomic drug delivery), while providing excellent anti-aggregation stability.[43] However, DSPE-PMPC-modified liposomes as previously studied are negatively charged and do not adsorb on negatively charged surfaces from water with no added salt. In contrast, such DSG-PMPC-modified liposomes are almost neutral and can adsorb on mica surfaces in pure water through dipole-charge interactions. Such DSG-PMPC-modified liposomes adsorb on negatively charged surfaces with providing superlubricating properties. It is appropriate to recall that while this study uses a negatively-charged mica surface as a model substrate, articular cartilage in synovial joints is also net-negatively charged,[55, 56]. We also note that, while we did not measure adsorption or lubrication by our novel, neutral DSG-PMPC modified HSPC liposomes in physiological-level salt solutions but only in pure water, we believe they would perform comparably well also at high salt. This is because, as shown earlier HSPC liposomes provide ultralow friction under physiologically high salt concentrations on mica surfaces.[57] At the same time, as demonstrated previously,[43] negatively-charged DSPE-PMPC stabilized HSPC liposomes are also very efficient lubricants when suitably screened at high salt, so that our neutral polyphosphocholinated vesicles would be expected to perform at least as well. Finally, we note that, quite apart from their boundary lubricating properties, the charge on functionalized liposomes may have a crucial effect on their use as drug delivery vehicles,[58] and here too our novel, neutral lipid-PMPC conjugates may play an important role.

**CRediT authorship contribution statement**

Weifeng Lin: Conceptualization, Methodology, Investigation, Writing – review & editing. Nir Kampf: Methodology, Investigation, Writing – review & editing. Jacob Klein: Conceptualization, Methodology, Writing – review & editing, Supervision, Funding acquisition.

**Declaration of Competing Interest**

Jacob Klein and Weifeng Lin have a patent on polyphosphocholinated-lipid conjugates and liposomes stabilized by such conjugates (US10730976B2).

**Acknowledgements**

We would like to thank the McCutchen Foundation and the Israel Science Foundation-National Natural Science Foundation of China Joint Program (Grants 2577/17 and 3618/21) for support of this work. This project has received funding from the European Research Council (ERC) under the European Union's Horizon 2020 research and innovation programme (Grant 743016). This work was made possible in part through the historic generosity of the Harold Perlman family.

# References

[1] C.-h. Huang, Phosphatidylcholine vesicles. Formation and physical characteristics, Biochemistry, 8 (1969) 344-352.

[2] L. Van Golde, J.J. Batenburg, B. Robertson, The pulmonary surfactant system: biochemical aspects and functional significance, Physiol. Rev., 68 (1988) 374-455.

[3] B. Hills, Remarkable anti-wear properties of joint surfactant, Ann. Biomed. Eng., 23 (1995) 112-115.

[4] L. Cwiklik, Tear film lipid layer: A molecular level view, Biochimica Et Biophysica Acta (BBA)-Biomembranes, 1858 (2016) 2421-2430.

[5] A. Sarma, G. Powell, M. LaBerge, Phospholipid composition of articular cartilage boundary lubricant, J. Orthop. Res., 19 (2001) 671-676.

[6] M.K. Kosinska, G. Liebisch, G. Lochnit, J. Wilhelm, H. Klein, U. Kaesser, G. Lasczkowski, M. Rickert, G. Schmitz, J. Steinmeyer, A lipidomic study of phospholipid classes and species in human synovial fluid, Arthritis. Rheum., 65 (2013) 2323-2333.

[7] W. Hodge, R. Fijan, K. Carlson, R. Burgess, W. Harris, R. Mann, Contact pressures in the human hip joint measured in vivo, Proc. Natl. Acad. Sci., 83 (1986) 2879-2883.

[8] W. Lin, J. Klein, Recent progress in cartilage lubrication, Adv. Mater., 33 (2021) 2005513.

[9] L. Grant, F. Tiberg, Normal and lateral forces between lipid covered solids in solution: correlation with layer packing and structure, Biophys. J., 82 (2002) 1373-1385.

[10] R. Goldberg, A. Schroeder, G. Silbert, K. Turjeman, Y. Barenholz, J. Klein, Boundary lubricants with exceptionally low friction coefficients based on 2D close‐packed phosphatidylcholine liposomes, Adv. Mater., 23 (2011) 3517-3521.

[11] M. Wang, C. Liu, E. Thormann, A. Dedinaite, Hyaluronan and phospholipid association in biolubrication, Biomacromolecules, 14 (2013) 4198-4206.

[12] Y. Duan, Y. Liu, J. Li, S. Feng, S. Wen, AFM study on superlubricity between Ti6Al4V/polymer surfaces achieved with liposomes, Biomacromolecules, 20 (2019) 1522-1529.

[13] A.-M. Trunfio-Sfarghiu, Y. Berthier, M.-H. Meurisse, J.-P. Rieu, Role of nanomechanical properties in the tribological performance of phospholipid biomimetic surfaces, Langmuir, 24 (2008) 8765-8771.

[14] T. Murakami, S. Yarimitsu, K. Nakashima, Y. Sawae, N. Sakai, Influence of synovia constituents on tribological behaviors of articular cartilage, Friction, 1 (2013) 150-162.

[15] W. Lin, M. Kluzek, N. Iuster, E. Shimoni, N. Kampf, R. Goldberg, J. Klein, Cartilage-inspired, lipid-based boundary-lubricated hydrogels, Science, 370 (2020) 335-338.

[16] P. Hilšer, A. Suchánková, K. Mendová, K.E. Filipič, M. Daniel, M. Vrbka, A new insight into more effective viscosupplementation based on the synergy of hyaluronic acid and phospholipids for cartilage friction reduction, Biotribology, 25 (2021) 100166.

[17] Y. Lei, Y. Wang, J. Shen, Z. Cai, C. Zhao, H. Chen, X. Luo, N. Hu, W. Cui, W. Huang, Injectable hydrogel microspheres with self-renewable hydration layers alleviate osteoarthritis, Sci. Adv., 8 (2022) eabl6449.

[18] S. Huang, B. Wang, X. Zhao, S. Li, X. Liang, R. Zeng, W. Li, X. Wang, Phospholipid reinforced P (AAm-co-AAc)/Fe3+ hydrogel with ultrahigh strength and superior tribological performance, Tribol. Int., (2022) 107436.
[19] J. Klein, Hydration lubrication, Friction, 1 (2013) 1-23.
[20] W.H. Briscoe, Aqueous boundary lubrication: Molecular mechanisms, design strategy, and terra incognita, Curr. Opin. Colloid Interface Sci., 27 (2017) 1-8.
[21] U. Raviv, J. Klein, Fluidity of bound hydration layers, Science, 297 (2002) 1540-1543.
[22] A. Adler, Y. Inoue, K.N. Ekdahl, T. Baba, K. Ishihara, B. Nilsson, Y. Teramura, Effect of liposome surface modification with water-soluble phospholipid polymer chain-conjugated lipids on interaction with human plasma proteins, J. Mater. Chem. B, 10 (2022) 2512-2522.
[23] D.D. Lasic, Sterically stabilized vesicles, Angew. Chem. Int. Ed., 33 (1994) 1685-1698.
[24] R. Tenchov, R. Bird, A.E. Curtze, Q. Zhou, Lipid Nanoparticles─ From Liposomes to mRNA Vaccine Delivery, a Landscape of Research Diversity and Advancement, ACS Nano, 15 (2021) 16982-17015.
[25] L.K. Müller, K. Landfester, Natural liposomes and synthetic polymeric structures for biomedical applications, Biochem. Biophys. Res. Commun., 468 (2015) 411-418.
[26] V. De Leo, F. Milano, A. Agostiano, L. Catucci, Recent advancements in polymer/liposome assembly for drug delivery: from surface modifications to hybrid vesicles, Polymers, 13 (2021) 1027.
[27] J.M. Harris, R.B. Chess, Effect of pegylation on pharmaceuticals, Nat. Rev. Drug Discovery, 2 (2003) 214-221.
[28] Z. Cao, L. Zhang, S. Jiang, Superhydrophilic zwitterionic polymers stabilize liposomes, Langmuir, 28 (2012) 11625-11632.
[29] Y. Li, Q. Cheng, Q. Jiang, Y. Huang, H. Liu, Y. Zhao, W. Cao, G. Ma, F. Dai, X. Liang, Enhanced endosomal/lysosomal escape by distearoyl phosphoethanolamine-polycarboxybetaine lipid for systemic delivery of siRNA, J. Control. Release, 176 (2014) 104-114.
[30] O.V. Zaborova, S.K. Filippov, P. Chytil, L. Kováčik, K. Ulbrich, A.A. Yaroslavov, T. Etrych, A novel approach to increase the stability of liposomal containers via in prep coating by poly [N‐(2‐hydroxypropyl) methacrylamide] with covalently attached cholesterol groups, Macromol. Chem. Phys., 219 (2018) 1700508.
[31] R. Hoogenboom, Poly (2‐oxazoline) s: a polymer class with numerous potential applications, Angew. Chem. Int. Ed., 48 (2009) 7978-7994.
[32] T. Ishihara, T. Maeda, H. Sakamoto, N. Takasaki, M. Shigyo, T. Ishida, H. Kiwada, Y. Mizushima, T. Mizushima, Evasion of the accelerated blood clearance phenomenon by coating of nanoparticles with various hydrophilic polymers, Biomacromolecules, 11 (2010) 2700-2706.
[33] M. Miyazaki, E. Yuba, H. Hayashi, A. Harada, K. Kono, Hyaluronic acid-based pH-sensitive polymer-modified liposomes for cell-specific intracellular drug delivery systems, Bioconjug. Chem., 29 (2018) 44-55.
[34] V.V. Khutoryanskiy, Beyond PEGylation: alternative surface-modification of nanoparticles with mucus-inert biomaterials, Adv. Drug Delivery Rev., 124 (2018) 140-149.

[35] M. Tully, M. Dimde, C. Weise, P. Pouyan, K. Licha, M. Schirner, R. Haag, Polyglycerol for half-life extension of proteins—alternative to PEGylation?, Biomacromolecules, 22 (2021) 1406-1416.

[36] C. Leng, H.-C. Hung, O.A. Sieggreen, Y. Li, S. Jiang, Z. Chen, Probing the surface hydration of nonfouling zwitterionic and poly (ethylene glycol) materials with isotopic dilution spectroscopy, J. Phys. Chem. C., 119 (2015) 8775-8780.

[37] S. Chen, L. Li, C. Zhao, J. Zheng, Surface hydration: Principles and applications toward low-fouling/nonfouling biomaterials, Polymer, 51 (2010) 5283-5293.

[38] Z. Cao, S. Jiang, Super-hydrophilic zwitterionic poly (carboxybetaine) and amphiphilic non-ionic poly (ethylene glycol) for stealth nanoparticles, Nano Today, 7 (2012) 404-413.

[39] J. Liu, J. Wang, Y.-f. Xue, T.-t. Chen, D.-n. Huang, Y.-x. Wang, K.-f. Ren, Y.-b. Wang, G.-s. Fu, J. Ji, Biodegradable phosphorylcholine copolymer for cardiovascular stent coating, J. Mater. Chem. B, 8 (2020) 5361-5368.

[40] Z. Wu, S. Li, Y. Cai, Y. Chen, X. Luo, Synergistic action of doxorubicin and 7-Ethyl-10-hydroxycamptothecin polyphosphorylcholine polymer prodrug, Colloids Surf. B, 189 (2020) 110741.

[41] X. Shi, D. Cantu-Crouch, V. Sharma, J. Pruitt, G. Yao, K. Fukazawa, J.Y. Wu, K. Ishihara, Surface characterization of a silicone hydrogel contact lens having bioinspired 2-methacryloyloxyethyl phosphorylcholine polymer layer in hydrated state, Colloids Surf. B, 199 (2021) 111539.

[42] J. Niu, H. Wang, J. Chen, X. Chen, X. Han, H. Liu, Bio-inspired zwitterionic copolymers for antifouling surface and oil-water separation, Colloids Surf. A, 626 (2021) 127016.

[43] W. Lin, N. Kampf, R. Goldberg, M.J. Driver, J. Klein, Poly-phosphocholinated liposomes form stable superlubrication vectors, Langmuir, 35 (2019) 6048-6054.

[44] W. Lin, R. Goldberg, J. Klein, Poly-phosphocholination of liposomes leads to highly-extended retention time in mice joints, J. Mater. Chem. B, 2022 (2022) 2820-2827.

[45] J. Du, Y. Tang, A.L. Lewis, S.P. Armes, pH-sensitive vesicles based on a biocompatible zwitterionic diblock copolymer, J. Am. Chem. Soc., 127 (2005) 17982-17983.

[46] J. Faivre, B.R. Shrestha, J. Burdynska, G. Xie, F. Moldovan, T. Delair, S. Benayoun, L. David, K. Matyjaszewski, X. Banquy, Wear protection without surface modification using a synergistic mixture of molecular brushes and linear polymers, ACS Nano, 11 (2017) 1762-1769.

[47] B. Bharatiya, G. Wang, S.E. Rogers, J.S. Pedersen, S. Mann, W.H. Briscoe, Mixed liposomes containing gram-positive bacteria lipids: Lipoteichoic acid (LTA) induced structural changes, Colloids Surf. B, 199 (2021) 111551.

[48] E. Kumacheva, J. Klein, Simple liquids confined to molecularly thin layers. II. Shear and frictional behavior of solidified films, J. Chem. Phys., 108 (1998) 7010-7022.

[49] J.N. Israelachvili, Intermolecular and surface forces, Academic press2011.

[50] K.L. Johnson, K.L. Johnson, Contact mechanics, Cambridge university press1987.

[51] P. Van Hoogevest, A. Wendel, The use of natural and synthetic phospholipids as pharmaceutical excipients, Eur. J. Lipid Sci. Technol., 116 (2014) 1088-1107.

[52] J.R. Yazdi, M. Tafaghodi, K. Sadri, M. Mashreghi, A.R. Nikpoor, S. Nikoofal-Sahlabadi, J. Chamani, R. Vakili, S.A. Moosavian, M.R. Jaafari, Folate targeted PEGylated liposomes for the oral delivery of insulin: In vitro and in vivo studies, Colloids Surf. B, 194 (2020) 111203.

[53] J.A. Kulkarni, S. Chen, Y.Y.C. Tam, Scalable Production of Lipid Nanoparticles Containing Amphotericin B, Langmuir, 37 (2021) 7312-7319.

[54] M. Kobayashi, Y. Terayama, M. Kikuchi, A. Takahara, Chain dimensions and surface characterization of superhydrophilic polymer brushes with zwitterion side groups, Soft Matter, 9 (2013) 5138-5148.

[55] M. Huber, S. Trattnig, F. Lintner, Anatomy, biochemistry, and physiology of articular cartilage, Invest. Radiol., 35 (2000) 573-580.

[56] J.A. Buckwalter, H.J. Mankin, A.J. Grodzinsky, Articular cartilage and osteoarthritis, Instr. Course Lect., 54 (2005) 465.

[57] R. Goldberg, A. Schroeder, Y. Barenholz, J. Klein, Interactions between adsorbed hydrogenated soy phosphatidylcholine (HSPC) vesicles at physiologically high pressures and salt concentrations, Biophys. J., 100 (2011) 2403-2411.

[58] L. Dézsi, T. Fülöp, T. Mészáros, G. Szénási, R. Urbanics, C. Vázsonyi, E. Őrfi, L. Rosivall, R. Nemes, R.J. Kok, Features of complement activation-related pseudoallergy to liposomes with different surface charge and PEGylation: comparison of the porcine and rat responses, J. Control. Release, 195 (2014) 2-10.